\documentclass[aps,twocolumn,prc,epsfig,showpacs]{revtex4}
\usepackage{graphicx}
\begin{document}
\title{Frame Dependence of  the Pair Contribution to the
Pion Electromagnetic Form Factor in a Light-Front 
Approach \footnote{To appear Brazilian Journal of Physics ({\bf BJP}) (2003).}}
\author{ J. P. B. C. de Melo$^a$, T. Frederico$^b$, E. Pace$^c$ and
G. Salm\`e$^d$ }
\affiliation{
$^a$  Instituto de F\'\i sica Te\'orica,
Universidade Estadual Paulista
01405-900, S\~ao Paulo, SP, Brazil \\
$^b$  Dep. de F\'\i sica, Instituto Tecnol\'ogico da
Aeron\'autica, Centro T\'ecnico Aeroespacial, 12.228-900, S\~ao
Jos\'e dos Campos,
S\~ao Paulo, Brazil \\
$^c$  Dipartimento di Fisica, Universit\`a di Roma "Tor Vergata"
and Istituto Nazionale di Fisica Nucleare, Sezione Tor Vergata,
Via della Ricerca
Scientifica 1, I-00133   Roma, Italy \\
$^d$  Istituto Nazionale di Fisica Nucleare, Sezione Roma I, P.le
A. Moro 2,I-00185   Roma, Italy \\  }
\date{\today}

\begin{abstract}
The frame dependence of the pair-term contribution
to the electromagnetic form factor of the pion  is studied
within the Light Front approach. A symmetric ansatz for the
pion Bethe-Salpeter amplitude with a pseudo scalar coupling of
the constituent to the pion field is used.
In this model, the pair term   vanishes for the Drell-Yan condition,
while it is dominant for momentum transfer along  the
light-front direction.
\end{abstract}

 \maketitle

\section{Introduction}

Within the Front Form dynamics\cite{dirac},
where the state of the system is defined at $x^+=t+z=0$,
if one uses the impulse approximation of the plus
component of the electromagnetic current ($j^+$) in the Drell-Yan frame to
calculate the form factors, the pair production from the incoming photon
(pair-term contribution) is in general suppressed by
light-front momentum conservation (see, e.g., \cite{FS}). This was seen in
schematic covariant models for spin-zero  composite systems
\cite{sawicki,pipach,Pacheco2002}.  However, even in the Drell-Yan frame
the pair term is present in $j^+$ for spin-one systems and is
necessary to keep the rotational properties of the matrix
element of the current\cite{pach98,ji2}.

To avoid the difficulties associated with the
rotational properties of the impulse approximation
some physically motivated schemes to
extract form factors from the current were used\cite{inna,to93,karmanov}.
In another approach, free of these ambiguities\cite{lev98},
the plus component of the momentum transfer is non zero while
the transverse momentum transfer vanishes in the Breit
frame. This can be achieved  by departing from the Drell-Yan
condition by rotating the system around the $y$-direction, i.e,
a non-kinematical transformation, and thus changing the direction
of the momentum transfer in the $z - x$ plane.

However, by relaxing the Drell-Yan condition, a light-front pair
term can contribute to the plus component of the  current, which
can be studied  in the pion example, as a prototype of a
relativistic system of bound constituents. In this work, the
composite system of a constituent quark and antiquark, is
described by an ansatz for the Bethe-Salpeter amplitude which is
nonconstant and symmetric  with a pseudo scalar coupling of the
constituent to the pion field. Our aim here is to discuss, within
that model, the magnitude of the pair contribution to the pion
electromagnetic form factor for momentum transfers in the $z-x$
plane in the Breit-frame, as has been done in
Ref.\cite{Pacheco2002}.

The work is organized as follows. In Sec. II, we present the
model of the  pion Bethe-Salpeter amplitude  and its
eletromagnetic current in  impulse approximation.
In Sec. III, we discuss the numerical results for
the pion electromagnetic form factor where
the separate contribution of the pair term is given.
We also present our summary in Sec. III.

\section{Pion Model and Electromagnetic Current}

The electromagnetic current of the pion
is calculated in impulse approximation, using a pseudoscalar
coupling between pion and quark fields, given by the effective
Lagrangian (see, e.g. \cite{tob92}):
\begin{equation}
{{\cal L}_I= - \imath g \vec\Phi \cdot \overline q \gamma^5 \vec
\tau q  \, } ,
\label{lain}
\end{equation}
where $g=m/f_\pi$ is the coupling constant from
the Goldberg-Treiman relation at the quark level,  and
$m$ is the mass of the constituents and $f_\pi$ the pion decay
constant.

The electromagnetic current of $\pi^+$ in
impulse approximation is build from the covariant
expression, which correspond to the  Feynman triangle diagram
(see, e.g., \cite{Bro69}):
\begin{eqnarray}
j^\mu&=&-\imath 2 e
\frac{m^2}{f^2_\pi} N_c\int \frac{d^4k}{(2\pi)^4}
\Lambda(k,P^{\prime})
\Lambda(k,P)  \nonumber \\
&\times & Tr[S(k)
\gamma^5 S(k-P^{\prime}) \gamma^\mu
S(k-P) \gamma^5 ]\ ,  \label{jmu}
\end{eqnarray}
where $\displaystyle S(p)=\frac{1}{\rlap\slash p-m+\imath \epsilon} \, $,
$N_c=3$ is the number of colors, $P^{\mu}$ and
$P^{\prime {\mu}}=P^{\mu}+q^{\mu}$ are the initial and
final momenta of the system,
$q^{\mu}$ is the momentum transfer and $k^{\mu}$
the spectator quark momentum.
The factor 2 stems from isospin algebra.

Our ansatz for the analytical form of the
vertex function describing the momentum part of
the coupling between the constituents and pion is:
\begin{equation}
\Lambda(k,P)=
\frac{C}{(k^2-m^2_{R} + \imath \epsilon)}+
\frac{C}{((P-k)^2-m^2_{R}+
\imath \epsilon)} \ ,
\label{vertex}
\end{equation}
where $m_R$ is the regulator parameter.
By imposing the charge normalization condition $F_{\pi}(q^2=0)=1$,
the constant $C$ is fixed.
This model satisfies current conservation $q\cdot j=0$\cite{Pacheco2002}.

We consider Breit frames, with the
momentum transfer $q^+ \ne 0$ and using
the light-front variables, i.e. $k^+=k^0+k^3 \ , k^-=k^0-k^3 \ ,
\vec k_\perp\equiv(k^1,k^2)$, one has
\begin{eqnarray}
q^+=-q^-=\sqrt{-q^2}\sin \alpha , \
q_x=\sqrt{-q^2}\cos \alpha , \  q_y=0,
\label{alpha}  \end{eqnarray}
and  $ q^2=q^+q^--(\vec q_\perp)^2 $.
The Drell-Yan frame with $q^+=0$ is recovered for $\alpha=0$,
while the $q^+=\sqrt{-q^2}$ condition
\cite{lev98} comes with $\alpha=90^o$.
 (The angle $\theta$ of Ref.\cite{bakker01} corresponds to $\alpha+90^o$).

The pion electromagnetic form factor is extracted from the
general covariant expression:
\begin{equation}
j^\mu = e (P^{\mu}+P^{\prime \mu}) F_\pi(q^2) \ ,
\label{full}
\end{equation}
evaluating the plus component of the current in Eq.(\ref{jmu}),
which  has two non vanishing contributions:
\cite{sawicki,Pacheco2002,bakker01,ji1}:
\begin{eqnarray}
F_\pi(q^2)=F^{(I)}_\pi(q^2,\alpha)+F^{(II)}_\pi(q^2,\alpha) \ .
\label{ffactor}
\end{eqnarray}
The integration over the interval of $0~\le~ k^+ \ <  \ P^+$
defines $F^{(I)}_\pi(q^2,\alpha)$ (see Fig. 1(a)).
For $k^+$ in the integration interval $P^+~\le~ k^+ ~ \le ~ P^{'+}$
(see Fig. 1(b)) one defines $F_\pi^{(II)}(q^2,\alpha)$.
In this model the contribution of the valence component of the
 wave function is part of
$F_\pi^{(I)}(q^2,\alpha)$. The pair term contribution to the  form
factor, with $q^+ \ > \ 0$, is  $F_\pi^{(II)}(q^2,\alpha)$.

\begin{figure}[thbp]
\centerline{\includegraphics[angle=0.0,width=7.0cm]{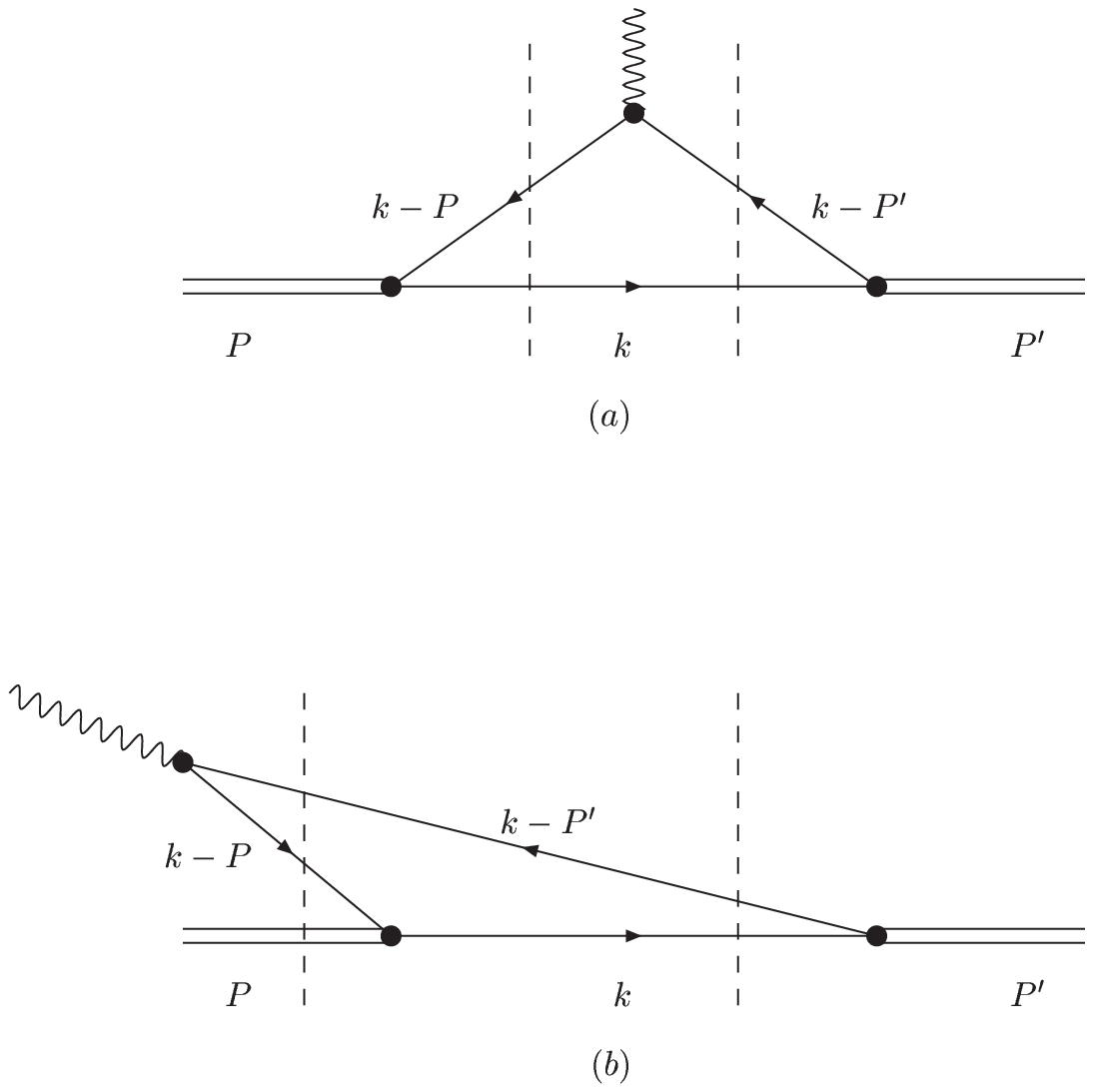}}
\caption[dummy0]{Light-front time-ordered diagrams for the current:
(a) $F^{(I)}_\pi$ (Eq.(\ref{FI}))  and (b) $F^{(II)}_\pi$ (Eq. (\ref{FII})).}
\label{fig1}
\end{figure}

The two contributions to the form factor are given by:
\begin{eqnarray}
F_\pi^{(I)}(q^2,\alpha)&=& - \imath
\frac{m^2}{(P^{+}+P^{\prime +})f^2_\pi}
\frac{N_c}{(2\pi)^4}
 \int d^{2} k_{\perp} d k^{+} d k^-
\nonumber \\ &\times &
\frac{\theta (k^+)\theta(P^+-k^+)}
{k^+(P^+-k^+) (P^{^{\prime}+}-k^+)}
\Pi(k,P,P')
\label{FI}
\end{eqnarray}
and
\begin{eqnarray}
F_\pi^{(II)}(q^2,\alpha)&=& - \imath
\frac{m^2 }{(P^{+}+P^{\prime +})f^2_\pi}
\frac{N_c}{(2\pi)^4}
 \int d^{2} k_{\perp} d k^{+} d k^-
\nonumber \\ &\times &
\frac{\theta (k^+-P^+)
\theta (P^{\prime +}-k^+)}
{k^+(P^+-k^+) (P^{^{\prime}+}-k^+)} \Pi(k,P,P')
\label{FII}
\end{eqnarray}
where
\begin{eqnarray}
&&\Pi (k,P,P')= \frac{Tr[{\cal O}^+]\Lambda(k,P) \Lambda(k,P^\prime)}
{
(P^- - k^- -(P-k)^-_{on}+ \frac{\imath\epsilon}{P^+ - k^+})}
\nonumber \\ &\times &
\frac{1}
{(k^- - k^-_{on}+\imath \epsilon)
(P^{\prime -} - k^- -(P^\prime-k)^-_{on}+\imath \epsilon)}
\ ,
\label{pi}
\end{eqnarray}
with,
\begin{eqnarray}
{\cal O}^+= (\rlap\slash k +m)\gamma^5(\rlap\slash k -
\rlap\slash P^\prime +m)
\gamma^+ (\rlap\slash k - \rlap\slash P + m) \gamma^5 \ .
\label{O}
\end{eqnarray}
The suffix $on$ indicates  particles on-$k^-$-shell.
In particular, for the Drell-Yan condition ($q^+=0$)
$F_\pi^{(II)}(q^2,\alpha)$ vanishes.

The Dirac propagator written in terms of light-front momenta
has two parts \cite{brodsky}:
\begin{eqnarray}
\frac{\rlap\slash{k}+m}{k^2-m^2+\imath \epsilon}=
\frac{\rlap\slash{k}_{on}+m}{k^+(k^--k^-_{on}+\frac{\imath \epsilon}{k^+})}
+\frac{\gamma^+}{2k^+} \ ,
\label{inst}
\end{eqnarray}
where $k^-_{on}=(k^2_\perp+m^2)/k^+$.
In the right-hand side of Eq.(\ref{inst}), the first term  is propagating
in the light-front time and the second one is instantaneous.
This second term contributes to both $F^{(I)}_\pi(q^2)$
 and $F^{(II)}_\pi(q^2)$,  due to the analytic structure of the
symmetric vertex function of Eq. (\ref{vertex}). The contribution
of the instantaneous term is of nonvalence nature because it is
left out in the definition of the valence wave function, as we
discuss below.

The pion Bethe-Salpeter amplitude within the model is given by:
\begin{eqnarray}
&&\Psi (k,P) =
\nonumber \\
&&{m \over
f_{\pi}}~\frac{\rlap\slash{k}+m}{k^2-m^2+\imath \epsilon}
\gamma^5 \Lambda (k,P)
\frac{\rlap\slash{k}-\rlap\slash{P}+m}{(k-P)^2-m^2+\imath \epsilon}
 \ ,
\label{bsa} \end{eqnarray} from which the momentum component of
the valence light-front wave function, $\Phi(k^+,\vec k_\perp;
P^+,\vec P_\perp)$\cite{Pacheco2002}, is derived eliminating the
relative time between the quarks after dropping the instantaneous
terms of the external Dirac propagators\cite{sales2}. Also, the
factors containing gamma matrices in the numerator and the phase
space factors $k^+$ and $(P^+-k^+)$ appearing in the denominator
are left out\cite{Pacheco2002}, and then one gets:
\begin{eqnarray}
&&\Phi(k^+,\vec k_\perp; P^+,\vec P_\perp)=
\left[\frac{{\cal N}}
{(1-x)(m^2_{\pi}-{\cal M}^2(m^2, m_R^2))} \right.
\nonumber \\
&&\left.
+\frac{{\cal N}}
{x(m^2_{\pi}-{\cal M}^2(m^2_R, m^2))} \right]
\frac{P^+}{m^2_\pi-M^2_{0}}
\ .
\label{wf2}
\end{eqnarray}
where ${\cal N}=\sqrt{N_c}~C~m/{f_\pi}$,
is a normalization factor
and
$x=k^+/P^+$, with $0 \ \le \ x \ \le \ 1$;
${\cal M}^2(m^2_a, m_b^2)= \frac{k^2_\perp+m_a^2}{x}+\frac{%
(P-k)^2_\perp+m^2_{b}}{1-x}-P^2_\perp \ ;$
and the square of the free mass is $M^2_0 ={\cal M}^2(m^2, m^2)$.
The light-front wave function, Eq.(\ref{wf2}),
is symmetric by the interchange of
quark and antiquark momenta, therefore it is not plagued by
the conceptual difficulties associated with the use of the nonsymmetric
regulator \cite{bakker01}.

The electromagnetic form factor evaluated in the
Breit frame using only the valence component is given by:
\begin{eqnarray}
&&F_\pi^{(WF)}(q^2,\alpha)= \frac{1}{2\pi^3(P^{\prime +}+P^+)}
\int d^{2} k_{\perp}\int_0^{P^+} d k^{+} \nonumber \\
&\times&
\left [ k^-_{on}P^+P^{\prime +}+\frac12 \vec k_\perp \cdot \vec q_\perp
(P^{\prime +}-P^{ +})-\frac14 k^+q^2_\perp \right ]
\nonumber \\
&\times&\frac{
\Phi(k^+,\vec k_\perp;P^{\prime +},\frac{\vec q_\perp}{2})
\Phi(k^+,\vec k_\perp;P^{ +},-\frac{\vec q_\perp}{2})
}{k^+(P^+-k^+) (P^{^{\prime}+}-k^+)}
\ ,
\label{Fwf}
\end{eqnarray}
where normalization constant $C$ is determined from the condition
$F_\pi(0)=1$ in Eq. (\ref{ffactor}). The probability of the
valence component in the pion, $\eta$, is identified to
$F_\pi^{(WF)}(0,0)$

The pion decay constant is one
constraint to fix the free parameters of the model:
\begin{eqnarray}
 P_\mu <0|A^\mu_i |\pi_j>= \imath~ m_\pi^2 f_\pi \delta_{ij} \ ,
\end{eqnarray}
where  $A^\mu_i = \bar{q} \gamma^\mu \gamma^5
\frac{\tau_i}{2} q$ is the isovector axial current.
With our ansatz for the
 pion-$\bar{q} q$ vertex function, one gets
\begin{eqnarray}
 f_\pi &=& -\imath\frac{m}{f_\pi} \frac{ N_c}{m^2_\pi}\int
\frac{d^4k}{(2\pi)^4} \nonumber \\
&\times & Tr[\rlap\slash P \gamma^5 S(k) \gamma^5 S(k-P)]
\Lambda(k,P)  \ ,  \label{f_pi}
\end{eqnarray}
and integrating on $k^-$, one arrives at
\begin{eqnarray}
f_{\pi} = \frac{m ~\sqrt{N_c}}{4\pi^3}
\int \frac{d^{2} k_{\perp} d k^+ } {k^+(m_\pi-k^+)}
\Phi(k^+,\vec k_\perp;m_\pi,\vec 0) \ ,
\label{fpi}
\end{eqnarray}
expressed in terms of the valence component of
the wave function \cite{tob92}.

\section{Numerical Results and Summary}

In this model, we have two free parameters:
the constituent quark mass $m$, which is chosen as 0.220 GeV
\cite{tob92,gi,inna94} and the regulator mass, $m_R$,
found to be 0.6 GeV from the fit of the experimental value
$f^{exp}_\pi=92.4$ MeV. The pion  mass used is 0.140 GeV.
With these parameters, the charge radius from
$\langle r^2 \rangle= 6\frac{\partial}{\partial q^2} F_\pi$,
comes out to be 0.74 fm, which is about $10 \%$ larger than
the experimental  value ($r_{exp}=0.67\pm 0.02$ fm \cite{amen}).
The probability of the $q\overline q$ Fock-state component
in the pion in the model is calculated to be $\eta=0.77$,
differently from  the nonsymmetric regulator model  of
Ref. \cite{pipach}, where $\eta=1$. We note that,
in a previous work\cite{tob92} it was necessary a
probability around 0.5 - 0.75 of the valence wave
function to fit the data on deep inelastic scattering.

\begin{figure}[hbtp]
\centerline{\includegraphics[angle=0.0,width=7.0cm]{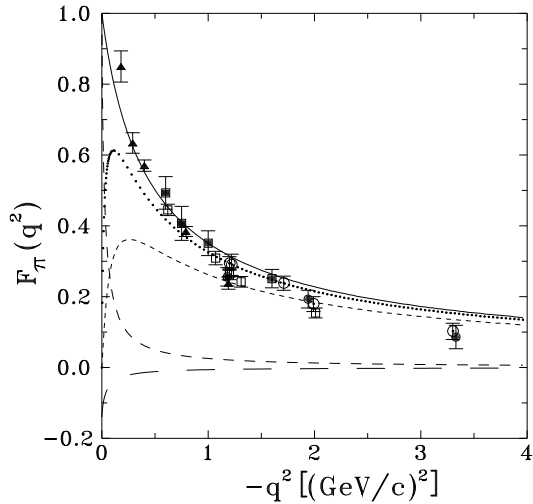}}
\caption[dummy0]{Pion form factor as a function of $ - q^2$ for
$\alpha=90^o$. Theoretical results: $F_\pi(q^2)$ (solid line);
$F_\pi^{(I)}(q^2,\alpha)$ (dashed line);
$F_\pi^{(II)}(q^2,\alpha)$ (dotted line); $F_{\pi \
inst}^{(I)}(q^2,\alpha)$
 (long-dashed line);  $F_{\pi \ inst}^{(II)}(q^2)$ (short-dashed line).
Experimental data: Ref. \cite{tj} (full squares),
 Ref. \cite{cea} (full triangles), Ref. \cite{corn1} (empty squares),
Ref. \cite{corn2} (empty circles) and Ref. \cite{bebek} (full circles).}
\label{fig2}
\end{figure}


In Fig. 2, the results for the
pion form factor  are shown and compared to the experimental data.
The full-model calculations, Eq. (\ref{ffactor}),
nicely agree with the new data for the pion form factor \cite{tj}.
Therefore, our symmetric vertex model can reproduce
the form factor data  consistently with the experimental
value of $f_\pi$, while for the nonsymmetric regulator
this was not achieved\cite{pipach}. We observe that the
model  reproduces simultaneously $f_\pi$ and the
experimental form factor for constituent quark mass in the
range between 0.2 and 0.3 GeV.

The separate contributions to the pion form factor,
$F_\pi^{(I)}$ and $F_\pi^{(II)}$ for $\alpha=90^o$,
are shown in Fig. 2. Differently from the case $\alpha=0^o$,
the form factor  is dominated by the pair production process
for $\alpha=90^o$, except near $q^2=0$.
Also, we observe that the form factor is completely
dominated by the pair-term contribution at high values
of the momentum transfer, which  appears to be fairly model
independent as well\cite{Pacheco2002,bakker01}.

\begin{figure}[thbp]
\centerline{\includegraphics[angle=0.0,width=7.0cm]{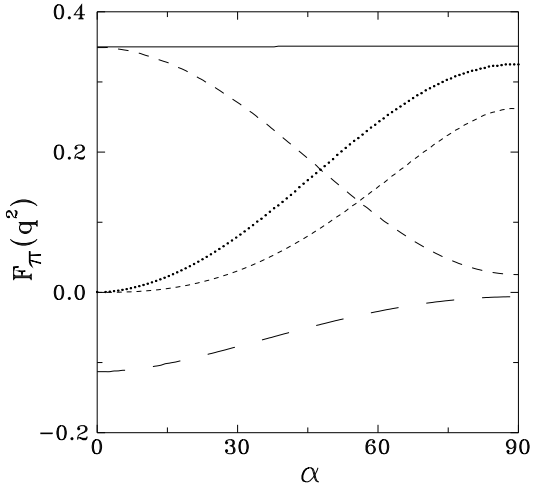}}
\caption[dummy0]{Contributions to the  pion form factor vs
$\alpha$ for $-q^2 =$ 1 (GeV/c)$^2$. Theoretical results:
$F_\pi(q^2)$ (solid line); $F_\pi^{(I)}(q^2,\alpha)$ (dashed line)
; $F_\pi^{(II)}(q^2,\alpha)$ (dotted line); $F_{\pi \
inst}^{(I)}(q^2,\alpha)$ (long-dashed line); $F_{\pi \
inst}^{(II)}(q^2,\alpha)$ (short-dashed line). } \label{fig3}
\end{figure}

The contributions of the instantaneous part of the Dirac
propagator to $F_\pi^{(I)}(q^2,\alpha)$ and
$F_\pi^{(II)}(q^2,\alpha)$,  called $F_{\pi \
inst}^{(I)}(q^2,\alpha)$ and  $F_{\pi \ inst}^{(II)}(q^2,\alpha)$,
respectively, are also shown in Fig. 2 for $\alpha=90^o$. The
value  of $F_{\pi \ inst}^{(I)}(q^2,\alpha)$ is nonzero because of
the specific analytic structure of the vertex
function\cite{Pacheco2002}. We also observe that $F_{\pi \
inst}^{(II)}(q^2,\alpha)$ dominates $F_{\pi}(q^2)$ at higher
momentum transfers. One can understand this by looking at the
diagram of Fig. 1(b), where in principle the spectator quark can
be exchanged between  the incoming and outgoing pion at a given
instant $x^+$, while the quark-antiquark  pair has been produced
by the virtual photon at an earlier stage. As the magnitude of the
momentum $q^-(=-q^+)$ increases, the time fluctuation for the
virtual process decreases and favors the instantaneous exchange of
the spectator quark between the initial and final pion, which
finally explains the dominance of $F_{\pi \
inst}^{(II)}(q^2,\alpha)$ in the pion form factor. In Fig. 3,  the
results for the various contributions to the pion form factor for
$-q^2=1$ (GeV/c)$^2$ as a function of the angle $\alpha$ are
shown. For increasing angles, the form factor changes smoothly
from valence to pair-term or nonvalence dominance.

In summary, we verified that the new data for the pion
electromagnetic form factor\cite{tj} is satisfactory described by
our symmetric ansatz of the vertex function  when the experimental
value of $f^{exp}_\pi$ is fitted   and the constituent quark mass
is chosen in the range between 0.2 and 0.3 GeV. We performed a
detailed analysis of the contribution of the light-front pair-term
to the form factor for different momentum transfer directions in
the Breit frame.  Such contribution is unique and does not depend
on the particular  choice of Bethe-Salpeter vertex, as long as the
four-dimensional impulse approximation is used to calculate the
electromagnetic current.  Another interesting outcome of our
symmetric model is  that the probability of the pion valence
component, $\eta$, is about 0.77), at variance with  previous
covariant calculations where  $\eta~=~1$\cite{pipach,bakker01}.

We thank the funding agencies FAPESP (Funda\c{c}\~{a}o de Amparo a
Pesquisa do Estado de S\~{a}o Paulo), and CNPq (Conselho Nacional
de Pesquisa e Desenvolvimento), and  Ministero della Ricerca
Scientifica e Tecnologica for partial support.


\begin{references}

\bibitem{dirac} P. A. M. Dirac, Rev. Mod. Phys. {\bf 21}, 392 (1949).

\bibitem{FS} L.L. Frankfurt and M.I. Strikman,
Nucl. Phys. {\bf B148}, 107 (1979).

\bibitem{sawicki}  M. Sawicki, Phys. Rev. {\bf D44}, 433(1991);
Phys. Rev. {\bf D46}, 474 (1992).

\bibitem{pipach}J. P. B. C. de Melo, H. W. Naus and T. Frederico,
 Phys. Rev.  {\bf C59}, 2278 (1999).

\bibitem{Pacheco2002}J. P. B. C. de Melo, T. Frederico,
E. Pace and G. Salm\`e,
Nucl. Phys. {\bf A707}, 399 (2002).


\bibitem{pach98}  J. P. B. C. de Melo, J. H. O. Sales, T. Frederico and
P. U. Sauer, Nucl. Phys. {\bf A631}, 574c (1998);
J. P. B. C. de Melo, T. Frederico, H. W. L. Naus and
P. U. Sauer, Nucl.Phys. {\bf A660}, 219 (1999).

\bibitem{ji2} B.L.G. Bakker, H.-M. Choi and C.-R. Ji,
Phys. Rev. {\bf D}, 116001 (2002).

\bibitem{inna}I. L. Grach and
L. A. Kondratyuk, Sov. J. Nucl. Phys. {\bf 39}, 198 (1984).

\bibitem{to93}L. L. Frankfurt, T. Frederico and M. I. Strikman,
Phys. Rev. {\bf C48}, 2182 (1993).

\bibitem{karmanov} J. Carbonell, B. Deplanques,
V. A. Karmanov and J.-F. Mathiot, Phys. Rep. {\bf 300}, 215 (1998).

\bibitem{lev98}F. M. Lev, E. Pace and G. Salm\`e,
Nucl. Phys. {\bf A641}, 229 (1998); Few-Body Syst. Suppl. {\bf
10}, 135 (1998); Phys. Rev. Lett. {\bf 83}, 5250 (1999); Phys.
Rev. {\bf C62}, 064004 (2000); Nucl. Phys. {\bf A663}, 365 (2000);
E. Pace and G. Salm\`e, Nucl. Phys. {\bf A684}, 487 (2001); {\bf
A689}, 411 (2001).

\bibitem{tob92}  T. Frederico and G. A. Miller, Phys. Rev.
{\bf D45}, 4207 (1992); Phys. Rev. {\bf D50}, 210 (1994).

\bibitem{Bro69} S.J. Brodsky and
J. R. Primack, Ann. Phys. {\bf 52}, 315 (1969), and references
therein quoted; F. Coester and  D.O. Riska, Ann. Phys. {\bf 234},
141 (1994).

\bibitem{bakker01}B. L. G. Bakker, H.-M. Choi
and C.-R. Ji, Phys. Rev. {\bf D63}, 074014 (2001).

\bibitem{ji1} H.-M. Choi and C.-R. Ji, Phys. Rev. {\bf D58}, 071901 (1998);
Phys. Rev. {\bf D59}, 034001 (1999).

\bibitem{brodsky}S. J. Brodsky, H. C. Pauli, and
S. S. Pinsky, Phys. Rep. {\bf 301}, 299 (1998).
  
\bibitem{sales2}J. H. O. Sales, T. Frederico,
B. V. Carlson and P. U. Sauer, Phys. Rev. {\bf C63}, 064003 (2001).

\bibitem{gi}S. Godfrey and N. Isgur, Phys. Rev. {\bf D32}, 185 (1985).

\bibitem{inna94}  F. Cardarelli, I.L. Grach, I.M.
Narodetskii, E. Pace, G. Salm\`e and S. Simula, Phys. Lett.
{\bf B332}, 1 (1994).

\bibitem{amen} S. R. Amendolia, G. Batignani, 
G. A. Beck, E.H. Bellamy, E. Bertolucci, 
G. Bologna, L. Bosito, C. Bradaschia, M. Budinich, 
M. Dell'Orso, 
B. D'Ettore Piazzoli, F.L. Fabbri, 
F. Fidecaro, L. Foa, E. Focardi, S.G.F. Frank, 
P. Gianetti, A. Giazzotto, M.A. Giorgi, M.G. Green, 
G.P. Heath, M.P.J. Landon, P. Laurelli, 
F. Liello, G. Mannocchi, 
P.V. March, P.S. Marrocchesi, A. Menzione, E. Meroni, 
P. Picchi, F. Ragusa, L. Ristori, L. Rolandi, A. Scribano, 
A. Stefani, D. Storey, J.A. Strong, R. Tenchini, G. Tonelli, 
G. Triggiani, W. Von Schlippe and A. Zallo, 
Phys. Lett. {\bf B178}, 116 (1986).

\bibitem{tj} J. Volmer, 
D. Abbott, H. Anklin, C. Armstrong, J. arrington, K. Assamagan, 
S. Avery, O. K. Baker, H.P. Blok, C. Bochna, 
E.J. Brash, H. Breuer, N. Chant, J. Dunne, T. Eden, 
R. Ent, D. Gaskell, R. Gilman, K. Gustafsson, 
W. Hinton, G.M. Huber, H. Jackson, M.K. Jones, 
C. Keppel, P.H. Kim, W. kim, 
A. Klein, D. Koltenuk, M. Liang, G.J. Lolos, 
A. Lung, D.J. Mack, D. MacKee, D. Meekins, 
J. Mitchel, H. Mkrcthyan, B. Muller, G. Niculescu, 
I. Niculescu, D. Pitz, D. Potterveld, 
L.M. Qin, J. Reinhold, I.K. Shin, S. Spepanyan, 
V. Tadevosyan, L.G. Tang, R.L.J. van der Meer, 
K. Vansyoc, D. Van Westrum, W. Vulcam, S. Wood, 
C. Yan, W.-X. Zhao and B. Zihlmann, Phys. Rev. Lett. {\bf 86}, 1713 (2001).

\bibitem{cea} C. N. Brown, 
C.R. Canizares, W.E. Cooper, A.M. Eisner, G.J. Feldman, 
C.A. Lichtenstein, L. Litt, W. Lockeretz, V.B. Montana and 
F.M. Pipkin, 
Phys. Rev. {\bf D8}, 92 (1973).

\bibitem{corn1}  
C. J. Bebek, C.N. Brown, M. Herzlinger, S. Holmes, 
C.A. Lichtenstein, F.M. Pipkin and L.K. Sisterson, 
Phys. Rev. {\bf D9}, 1229 (1974).

\bibitem{corn2}  C. J. Bebek, 
C.N. Brown, M. Herzlinger, S.D. Holmes, F.M. Pipkin, 
S. Raither and L.K. Sisterson, 
Phys. Rev. {\bf D13}, 25 (1976).

\bibitem{bebek}  C. J. Bebek, 
C.N. Brown, S.D. Holmes, R.V. Kline, F.M. Pipkin, 
S. Raither and L.K. Sisterson, 
Phys. Rev. {\bf D17}, 1693 (1978).

\end{references}
\end{document}